# Asymmetric longitudinal optical binding force between two identical dielectric particles with electric and magnetic dipolar responses

Xiao-Yong Duan,[1,2*] Graham D. Bruce,[2] Feng Li,[1] Kishan Dholakia,[2,3,4]

[1] *School of Data Science, Jiaxing University, Jiaxing 314001, China*
[2] *SUPA School of Physics and Astronomy, University of St Andrews, St Andrews, KY16 9SS, UK*
[3] *Department of Physics, College of Science, Yonsei University, Seoul 03722, South Korea*
[4] *School of Biological Sciences, The University of Adelaide, Adelaide, South Australia, Australia*

*Corresponding author: xyduan@zjxu.edu.cn

**Abstract:** In general, the optical binding force between identical particles is thought to be symmetric. However, we demonstrate analytically a counter-intuitively asymmetric longitudinal optical binding force between two identical electric and magnetic dipolar dielectric particles. This homodimer is confined in two counter-propagating incoherent plane waves along the dimer's axis. The force consists of the electric dipolar, magnetic dipolar, and electric-magnetic dipolar coupling interactions. The combined effect of these interactions is markedly different than the expected behavior in the Rayleigh approximation. The asymmetric force is a result of the asymmetric forward and backward scattering of the particles due to the dipolar hybridization and coupling interactions. Consequently, it leads to a harmonic driving force on the pair, which decays with the interparticle distance to the first power. We show the rich nonequilibrium dynamics of the dimer and of the two particles impelled by the driving and binding forces and discuss the ranges of particle refractive index and size in which the asymmetric binding force arises. Our results open perspectives for nonequilibrium light-driven multiparticle transport and self-assembly.

## I. INTRODUCTION

The optical binding force is mediated by the momentum exchange between light and matter to realize stable equilibrium configurations of tiny objects [1, 2]. Therefore, it is one of the most important tools in light-induced micro/nanoparticle self-assembly of, e.g., arrays [3, 4], metamolecule [5], colloid optical waveguide [6], and optical matter [2, 7]. In general, there are two forms of the force. One is the transverse optical binding force (TOBF) where the incident wave vector is perpendicular to the connecting-line of the two particles [8]. The other is termed the longitudinal optical binding force (LOBF), as the external light is incident on the two particles along the dimer's axis [9-11].

Recently, there is increasing interest in the nonreciprocal optical binding force between dissimilar particles [12]. The reason is that the symmetry-breaking of the action-reaction results in rich individual and collective nonequilibrium dynamics of the particles [13]. Therefore, it has potential application in light-driven nanomotors [14, 15] and nanoswimmers [16] as well as in the nonequilibrium assembly of colloids [17]. In detail, the TOBF between two dissimilar particles has been demonstrated in theory [18] and experiment [19] to be asymmetric in the Rayleigh approximation (where the particle size is much smaller than the incident wavelength). As a result, this causes a net driving force on the center of the heterodimer [20, 21]. Interestingly, the TOBF between two identical silver particles confined in an optical ring trap with an azimuthal phase gradient is also asymmetric [22]. Excitingly, the nonreciprocal TOBF was recently harnessed to realize an optical micro-gear capable of doing work: binding a probe particle around a hexagonal array composed of seven identical particles who are differently sized than the probe particle, and transferring spin angular momentum from the array to orbital angular momentum of the probe particle [23]. On the other hand, the numerical results of the coupled dipole method show that the asymmetric LOBF between two unequal polystyrene particles causes not only the motion of the center of the heterodimer but also a stable bonding structure [24, 25].

Traditionally, the optical binding force between two identical particles is symmetric [9], and there is therefore no net driving force on the pair [26]. In this work, however, we demonstrate analytically a counter-intuitively asymmetric LOBF for two identical dielectric particles with electric and magnetic dipolar responses. It is the result of the electric-magnetic dipolar hybridization and coupling interactions of the particles. Moreover, the resultant driving force is harmonic while its envelope is inversely proportional to the interparticle distance. Unexpectedly, the stable and unstable equilibrium positions of the driving force interchange depending on the particle size and interparticle distance.

The paper is organized as follows. In Sec. II, we present the analytical asymmetric LOBF and driving force. Moreover, their physical origins are revealed. In Sec. III, both the strengths and directions of the LOBF and driving force as well as the stable and unstable equilibrium positions of the two forces are numerically investigated in detail. In Sec. IV, the refractive index and size ranges of particle in which the asymmetric LOBF appears are discussed. Finally, conclusions are drawn in Sec. V.

## II. THEORETICAL MODEL

Here, two counter-propagating incoherent plane waves have been employed to trap two identical electric and magnetic dipolar dielectric particles along the homodimer's axis ($y$ axis) as shown in Fig. 1 (a). The mutual incoherence avoids the influence of the interference between two incident waves on the binding force [27]. In particular, the LOBF is polarization-independent because of the symmetry between the induced electric dipole (ED) pair and magnetic dipole (MD) pair in the dimer. Hence, we take $p$-polarized waves whose electric fields are along the $z$ axis as an example. In the framework of the optical force on a single dielectric particle [28] and electromagnetic wave mutual scattering [29] (or see Appendices in Ref. [30]), the LOBF is analytically decomposed into the electric dipolar, magnetic dipolar, and electric-magnetic dipolar coupling components. In detail, the electric dipolar component of the LOBF on particle j (j=A or B) is expressed as

$$F_e^j = \frac{2n_s I_0}{c} \mathrm{Re}\left[\begin{array}{c} \pm(\alpha_e k)\sin(kR)(\mu\alpha_e \mp i\eta\alpha_m) \\ \mp\frac{\partial \mu^*}{\partial R}|\alpha_e|^2 \cos(kR) \mp \frac{\partial \eta^*}{\partial R}\alpha_e \alpha_m^* \sin(kR) \end{array}\right], \quad (1)$$

where $I_0 = \varepsilon_0 n_s c |E_0|^2/2$ is the intensity of the incident wave in the medium with refractive index $n_s$, $\varepsilon_0$ is the permittivity of vacuum, $c$ is light speed in vacuum, $E_0$ is the amplitude of the incident electric field. The time-dependent factor of the incident wave is $\exp(-i\omega t)$ while $\omega$ is the angular frequency. $\alpha_e = i6\pi a_1/k^3$ and $\alpha_m = i6\pi b_1/k^3$ represent individually the electric and magnetic polarizabilities with radiation reaction terms [31] of the particles. $a_1$ and $b_1$ denote respectively the electric and magnetic dipolar Mie scattering coefficients [32]. $i$ is the unit imaginary number, Re represents the real part of a complex number, $^*$denotes complex conjugation. $R$ is the separation between the centers of the two particles [see Fig. 1 (a)], $\partial/\partial R$ denotes the partial derivative with respect to $R$. $k=2\pi/\lambda$ is the wavenumber in the medium, $\lambda$ and $\lambda_0$ are respectively the incident wavelength in the medium and vacuum with relation $\lambda=\lambda_0/n_s$. $\mu=\exp(ikR)(k^2R^2+ikR-1)/(4\pi R^3)$ and $\eta=\exp(ikR)(ik^2R^2-kR)/(4\pi R^3)$ are respectively the eigenvalues of the electric and magnetic dyadic Green's functions of a point dipole [30]. Note that the terms on the right side of Eq. (1) take the upper signs in the case of j=A while they revert to the lower signs for j=B. [The rule of the signs is also suitable for Eq. (2) .]

Consider the LOBF acting on particle B as an example. The first two terms in Eq. (1) represent the interaction between the incident electric field and the ED in particle B. The ED is induced by the radiated electric fields of the ED and MD in particle A. The third term shows the interaction between the radiated electric field of the ED in particle A and the ED in particle B which is caused by the incident electric field. The fourth term expresses the interaction between the radiated electric field of the MD in particle A and the ED in particle B, which is caused by the incident electric field. Note that the secondary interactions of the radiated fields by both ED and MD in particle A with the secondary ED and MD in particle B induced by these fields are in general very small [9]. Hence, they can be ignored within our parameter range comparing with the primary interactions in Eq. (1) [33]. On the other hand, the magnetic dipolar components of the LOBF ($F_m^j$) are also expressed by Eq. (1) through exchanging the subscript "e" and "m". In other words, the difference between $F_e$ and $F_m$ is just the $\alpha_e$ and $\alpha_m$ ($a_1$ and $b_1$) of the particles. Analogy with $F_e^j$, $F_m^j$ represents four interactions between the corresponding magnetic field and MD. Finally, the electric-magnetic dipolar coupling components of the LOBF are described by

$$F_{em}^j = \frac{n_s I_0}{3\pi c} k^4 \sin(kR) \mathrm{Im}\left[\alpha_e \alpha_m^* \begin{pmatrix} \mp\mu^*\alpha_m^* \pm \mu\alpha_e \\ \mp i(\eta^*\alpha_e^* \pm \eta\alpha_m) \end{pmatrix}\right], \quad (2)$$

where Im represents the imaginary part of a complex number.

Amazingly, the LOBF expressed by Eqs. (1) and (2) is nonreciprocal, as denoted by the second term $\alpha_e k i\eta\alpha_m$ in Eq. (1) and the fourth term $\alpha_e \alpha_m^* i\eta\alpha_m$ in Eq. (2). This is the first important conclusion in this paper. The underlying physics is the asymmetric forward and backward scattering of the homodimer on the $y$ axis. For the sake of clarity, we use the term forward (backward) scattering to refer to the scattering along the $y$ (-$y$) axis, i.e. to the right (left) of the dimer, as shown in Fig. 1 (a), in the two counter-propagating plane waves configuration. This asymmetric scattering is a result of the symmetry-breaking of the ED pair and MD pair [22], which are caused by different phases of the dipole moments, as expressed by

$$p^j = \varepsilon_0 \varepsilon_s \alpha_e \delta_e^j E_0, \quad (3)$$

$$m^j = \alpha_m \delta_m^j H_0, \quad (4)$$

where $H_0$ is the amplitude of the incident magnetic field and $\varepsilon_s$ is the permittivity of the medium. The phases of the moments are caused by both the scattering properties ($a_1$ and $b_1$) of particles and the relaxation of electromagnetic field between the two particles due to the phase shift $\Delta\varphi=kR$. The electric and magnetic relaxations are respectively determined by

$$\delta_e^j = 1 + 2\sin(kR)(\mu\alpha_e \mp i\eta\alpha_m), \quad (5)$$

$$\delta_m^j = 1 + 2\sin(kR)(\mu\alpha_m \mp i\eta\alpha_e). \quad (6)$$

The relaxation includes not only the incident field but also the dipolar hybridization and coupling interactions. In detail, the hybridization refers to the fact that the ED (MD) in one particle is affected by the radiated electric (magnetic) fields of the ED and MD in the neighboring particle [34]. On the other hand, the coupling comes from the interaction between the ED and MD in one particle [30]. For the ED (MD) moment in Eq. (3)[Eq. (4)] of one particle, Eq. (5) [Eq. (6)] denotes that the first term depends on the incident electric (magnetic) field while the last two terms are determined by the radiative electric (magnetic) field of the ED and MD in the other particle. Note that the last terms in the second brackets in Eqs. (5) and (6) take respectively the upper and lower signs for particles A and B.

To illustrate the asymmetric ED and MD pairs in a dielectric homodimer, we employ two identical polystyrene particles with refractive index $n_p$=1.59 and radius $a$=130nm immersed in air ($n_s$=1). The incident wavelength is $\lambda$=532nm throughout. Considering only the phases and relative intensities of the dipole moments, Eqs. (3) and (4) are simplified as the dimensionless $p^j = ia_1\delta_e^j$ and $m^j = ib_1\delta_m^j$ which are presented in Fig. 1 (b). It is clearly seen that $p^A$ (blue solid curve) and $p^B$ (blue double dot-dashed curve) are asymmetrical (nonidentical) in general, as are $m^A$ (red dashed curve) and $m^B$ (red dotted curve), with respect to $R$. The reason is that the $\Delta\varphi$ (top horizontal axis) causes nonzero phase difference $\sin(kR)\eta$ between the two EDs/MDs [see Eqs. (5) and (6)]. Therefore, these result in asymmetric forward and backward scattering of the homodimer [22] even in the symmetric geometry configuration seen in Fig. 1 (a). As expected, the two ED moments as well as the two MD moments become symmetrical (identical) at the positions $R=l\lambda/2$ ($l$ is integer number), i.e., $\lambda$ and $1.5\lambda$. These are determined by the zero phase difference which are only caused by $\sin(kR)=0$ due to $\Delta\varphi=kR=l\pi$, i.e., $\Delta\varphi=2\pi$ and $3\pi$, regardless of $\eta$. These result in symmetric forward and backward scattering of the homodimer. However, the two ED moments are also identical as well as the two MD moments at positions $R=1.1\lambda$ and $1.6\lambda$ in spite of $\Delta\varphi\neq l\pi$. The reason is that the phase difference is generally determined by not only $\sin(kR)$ but also $\eta$. The solution of $\sin(kR)\eta=0$ with $kR\neq l\pi$ is $kR=\pi/4+l\pi$. Therefore, the separations $R=1.1\lambda$ and $1.6\lambda$ correspond almost to the zero points of $\sin(kR)\eta$ when $l=2$ and 3. Interestingly, the symmetric ED pair and MD pair at these positions lead to asymmetric forward and backward scattering of the dimer, which will be seen in Fig. 3 (a).

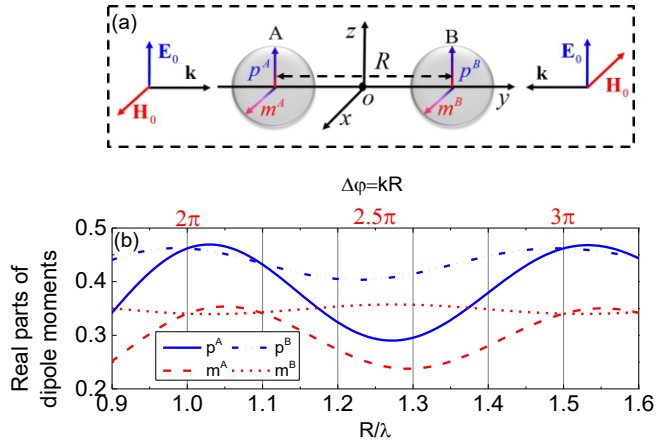

Fig. 1. Longitudinal optical binding between two identical electric and magnetic dipolar dielectric nanospheres with separation ($R$) immersed in air and illuminated by two incoherent counter-propagating plane waves. (a) Two $p$-polarized plane waves with electric $\mathbf{E}_0$ (magnetic $\mathbf{H}_0$) fields along the $z$ ($x$) axis propagate along the $y$ axis in Cartesian system ($O$-$xyz$) as shown by the wave vector $\mathbf{k}$. $p^j$ ($m^j$) denotes the induced ED (MD) moment in the two particles A and B along the $z$ ($x$) axis. (b) The real parts of the dimensionless $p^A$ (blue solid curve), $p^B$ (blue double dot-dashed curve), $m^A$ (red dashed curve), and $m^B$ (red dotted curve) as functions of $R$ (bottom horizontal axis) and $\Delta\varphi=kR$ (top horizontal axis).

Furthermore, the nonreciprocal LOBF causes a net driving force on the center of the homodimer. It is the sum of the LOBFs on the two particles and consists of the electric dipolar, magnetic dipolar, and electric-magnetic dipolar coupling components as

$$F_{\text{driv}} = F_{\text{driv}}^{\text{e}} + F_{\text{driv}}^{\text{m}} + F_{\text{driv}}^{\text{em}}, \quad (7)$$

where

$$F_{\text{driv}}^{\text{e}} = F_{\text{e}}^{\text{A}} + F_{\text{e}}^{\text{B}} = F_{\text{driv}}^{\text{m}} = \frac{4n_s I_0}{c} k \sin(kR) \text{Im}[\alpha_e \alpha_m \eta], \quad (8)$$

$$F_{\text{driv}}^{\text{em}} = F_{\text{em}}^{\text{A}} + F_{\text{em}}^{\text{B}} = -\frac{n_s I_0}{3\pi c} k^4 |\alpha_m|^2 \sin(kR) \text{Re}[\alpha_e \eta]. \quad (9)$$

In the case of large separation ($R>\lambda$), where only the highest order term of $kR$ in $\eta$ is retained because of $kR\gg 1$, Eqs. (8) and (9) are simplified as

$$F_{\text{driv}}^{\text{e}} = F_{\text{driv}}^{\text{m}} = \frac{18\pi n_s I_0}{ck^3 R} \left\{ \begin{array}{l} -\sin(2kR)\text{Re}[a_1 b_1] \\ +(1-\cos(2kR))\text{Im}[a_1 b_1] \end{array} \right\}, \quad (10)$$

$$F_{\text{driv}}^{\text{em}} = \frac{18\pi n_s I_0}{ck^3 R} |b_1|^2 \left\{ \begin{array}{l} \sin(2kR)\text{Re}[a_1] \\ -(1-\cos(2kR))\text{Im}[a_1] \end{array} \right\}. \quad (11)$$

Equations (10) and (11) demonstrate clearly that the driving force is harmonic as well as that its envelope is inversely proportional to the separation $R$. This is the second important conclusion in this paper. As expected, the asymmetric LOBF described by Eqs. (1) and (2) reduces to the symmetric one in the Rayleigh approximation ($\alpha_m=0$ or $b_1=0$) expressed by Eq. (25) in Ref. [9].

## III. RESULTS

The power density of the incident wave is $I_0=10\text{mW}/\mu\text{m}^2$. The force, radius, and the separation are respectively described in units of $p$N, nm, and $\lambda$ throughout the paper.

### A. Polystyrene homodimer

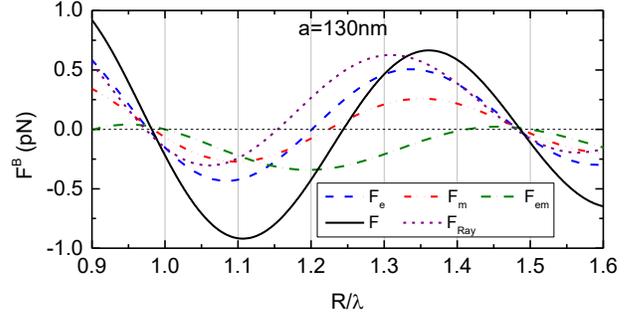

Fig. 2. LOBF $F^B$ (black solid curve) on polystyrene ($n_p=1.59$) sphere B (Fig. 1) with radius $a=130$nm as well as its electric dipolar $F_e$ (blue dashed curve), magnetic dipolar $F_m$ (red dash-dot-dotted curve), and electric-magnetic dipolar coupling $F_{em}$ (green dash-dotted curve) components as a function of separation ($R$). $F_{Ray}$ represents the classical LOBF (magenta short dashed curve) in Rayleigh approximation.

Consider particle B (Fig. 1) as an example. Figure 2 shows the LOBF $F$ (black solid curve) and its electric dipolar $F_e$ (blue dashed curve), magnetic dipolar $F_m$ (red dash-dot-dotted curve), and electric-magnetic dipolar coupling $F_{em}$ (green dash-dotted curve) components on the particle with respect to separation $R$. The results denote the remarkable contributions of $F_m$ and $F_{em}$ to LOBF even for low-refractive-index polystyrene ($n_p=1.59$) particles, which have been widely employed in optical binding experiments [35, 36], with radius $a=130$nm. In detail, the magnitude of the LOBF is largely enhanced due to the magnetic and coupling interactions comparing with the classical LOBF $F_{Ray}$ (magenta short dashed curve) calculated by Eq. (30) in Ref. [27]. The reason is that both EDs and MDs in the particles are effectively excited when $a\geqslant 130$nm even though they don't resonate [see Fig. 8 (a) in Appendix A]. Consequently, $F_{Ray}$ grossly underestimates the LOBF and even departs from $F_e$ where the magnetic and electric-magnetic coupling effects are considered. This is similar to the TOBF for dielectric particles [30]. On the other hand, we focus on the stable and unstable equilibrium positions of $F_{Ray}$ and LOBF (also termed the stable and unstable equilibrium separations of the two particles). At these positions, the force is zero and the slope of the force is respectively negative or positive. The stable equilibrium positions of $F_{Ray}$ deviate slightly from those of the LOBF near $R=\lambda$ and $1.5\lambda$. Furthermore, the unstable one of $F_{Ray}$ at $R=1.15\lambda$ deviates largely from the counterpart of LOBF at $R=1.25\lambda$. Of course, these phenomena are similar to the LOBF on particle A.

Figure 3 (a)-3 (c) show the LOBFs $F^A$ (red dashed curve) and $F^B$ (blue short dashed curve) on polystyrene particles A and B as well as the driving force $F_{driv}$ (black solid curve), which is the vector sum of $F^A$ and $F^B$, on the center of the homodimer for three different radii. $F^A$ and $F^B$ show that the LOBF is nonreciprocal as predicted by Eqs. (1) and (2). As a result, this leads to a net driving force $F_{driv}$ which is similar to that previously demonstrated for mismatched particles [24]. Importantly, we demonstrate analytically that $F_{driv}$ oscillates harmonically and decays with $R$ to the first power as shown by Eqs. (10) and (11). In addition, the electric and magnetic components $F_{driv}^e + F_{driv}^m$ (purple solid curve with square) as well as the coupling component $F_{driv}^{em}$ (orange solid curve with triangle) of $F_{driv}$ are presented in Fig. 3 (a). Note that $F_{driv}$ vanishes when $R=l\lambda/2$, i.e., $\lambda$ and $1.5\lambda$, because of $\sin(2kR)=1-\cos(2kR)=0$ regardless of the scattering properties ($a_1$ and $b_1$) in Eqs. (10) and (11). These separations correspond exactly to $\Delta\varphi=l\pi$ in Fig. 1 (b). Thus, the symmetric ED pair and MD pair cause symmetric forward and backward scattering which results in a reciprocal LOBF. Interestingly, the other separation $R=1.13\lambda$ where $F_{driv}=0$ does not coincide with the position $R=1.1\lambda$ in Fig. 1(b). The reason is that the zero phase difference $\sin(kR)\eta=0$ with $kR\neq h\pi$ results in a symmetric ED pair and MD pair when $R=1.1\lambda$. Then, the symmetric ED pair and MD pair result in zero $F_{driv}^e$ and $F_{driv}^m$, as shown by Eq. (10) and the purple curve with square in Fig. 3 (a), even though each term in Eq. (10) is nonzero. However, $F_{driv}^{em}$ (orange curve with triangle) is nonzero and leads to a net driving force at $R=1.1\lambda$. This is caused by the asymmetric coupling interactions between the two particles because of the different phase differences between the ED and MD in different particles due to the nonzero $\Delta\varphi$. Furthermore, the nonzero three components of $F_{driv}$ cancel each other and result in a zero $F_{driv}$ at $R=1.13\lambda$. A similar phenomenon also arises in Fig. 3 (c) as shown by the black dashed arrow. To illuminate the origination of $F_{driv}$ as shown by Eqs. (3)-(6), the far-field scattering patterns of the homodimer, wherein the observational point from the center of the two particles is much larger than $R$, in x-y (blue solid curves) and y-z (red dash curves) planes are presented in Fig. 3 (d). In detail, for 130nm particles with $R=1.05\lambda(1.3\lambda)$, the backward scattering is stronger (weaker) than the forward scattering. The result corresponds to the positive (negative) driving force in Fig. 3 (a). Similarly, for 163nm particles with $R=1.15\lambda$ and 172nm particles with $R=1.2\lambda$, the forward scattering is dominant over the backward scattering. They are consistent with the negative driving forces in Figs. 3 (b) and 3(c). Interestingly, $F_{driv}$ and LOBF induce complex dynamics of the dimer and of the two constituents, respectively. In detail, the stable equilibrium separations of the two particles (green solid arrows) are slightly shorter than the integer multiples of half a wavelength, i.e., they are found to be $0.98\lambda$ and $1.48\lambda$ rather than the expected $\lambda$ and $1.5\lambda$ respectively. Meanwhile, the unstable one (green dashed arrow) lies halfway between the two stable separations. With the unstable equilibrium separation, $F^A$ and $F^B$ are identical in magnitude and direction. Thus, the two particles move together along the $-y$ axis. But their relative motion is unstable and easily broken by any external force. Notice that the stable and unstable separations are independent of the particle size as shown in Fig. 3(a)-3(c). On the other hand, the unstable equilibrium positions of $F_{driv}$ (black dashed arrows) exactly correspond to $R=l\lambda/2$ ($l$ is positive integer), i.e., $R=\lambda$ and $1.5\lambda$, when $a=130$nm in Fig. 3(a). It means that the dimer's center is unstable even though the two particles have a stable separation at the positions. The black solid arrow shows the stable equilibrium position of $F_{driv}$ where the two particles attract each other but their common center is stable. In particular, the positive $F_{driv}$ pushes the dimer along the $y$ axis while the two particles are attractive. On the contrary, the negative $F_{driv}$ push the dimer along the $-y$ axis while the two particles are either attractive or repulsive.

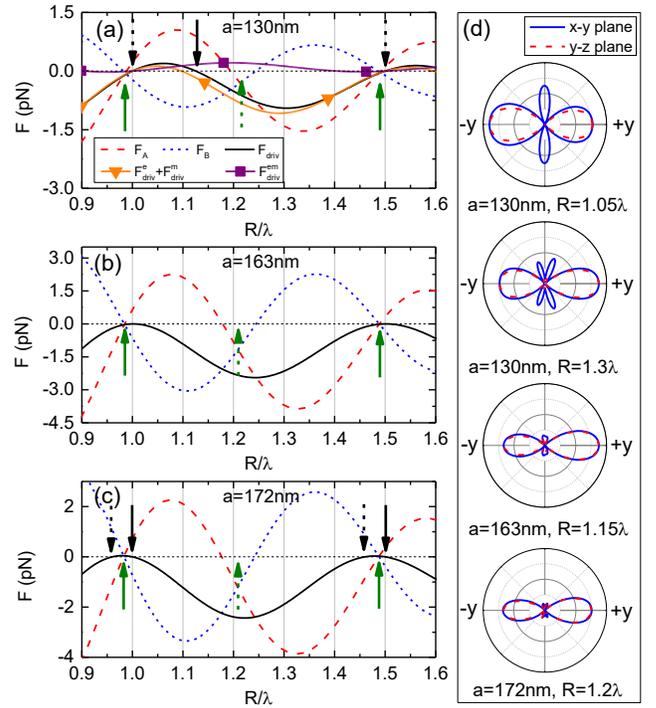

Fig. 3. LOBFs $F^A$ (red dashed curve) and $F^B$ (blue short dashed curve) on polystyrene particles A and B with radius $a=130$nm (a), 163nm (b), and 172nm (c) as well as the driving force $F_{driv}$ (black solid curve) on the center of the dimer as a function of separation ($R$). For 130nm particle in subplot (a), the electric and magnetic components $F_{driv}^e + F_{driv}^m$ (purple solid curve with square) as well as the coupling component $F_{driv}^{em}$ (orange solid curve with triangle) of $F_{driv}$ are presented. The black solid (dashed) arrows denote the stable (unstable) equilibrium positions of $F_{driv}$ while the green counterparts denote the stable and unstable separations between the two particles. (d) The four far-field scattering patterns of the homodimer in x-y (blue solid curves) and y-z (red dash curves) planes (see Fig. 1) correspond respectively to the dimer with different radii and separations.

Figure 4 shows $F_{driv}$ on a polystyrene dimer as well as its stable (back lines) and unstable (white lines) equilibrium positions and their dependence of the particle radius ($a$) and separation ($R$). The reason for choosing the range of particle radius from 100nm to

172nm is the particle exhibits strong electric and magnetic dipolar responses. It corresponds to the electric and magnetic dipolar model in this paper. On the contrary, the symmetric forward and backward scattering of the Rayleigh particle ($a$<100nm) with only ED does not cause net $F_{driv}$. On the other hand, larger particles ($a$>172nm) with higher multiple moments are beyond our model range [see Fig. 8 (a) in Appendix A]. The unstable equilibrium positions are located at $R=l\lambda/2$, i.e., $\lambda$, 1.5$\lambda$, and 2$\lambda$, when $a$<163nm. It is the result of $sin(2kR)=1-cos(2kR)=0$ in Eqs. (10) and (11). Further, the stable equilibrium positions are decreasing and getting closer and closer to the unstable ones with increase of the radius of particle. Interestingly, for $a$=163nm, the stable and unstable equilibrium positions are merged into one non-equilibrium zero-force position because $F_{driv}$ is non-positive as shown in Fig. 3 (b). It means that the two particles always move together along the -$y$ axis regardless of their separation and relative stability. Unexpectedly, the stable and unstable equilibrium positions exchange when the radius goes beyond 163nm, as also shown by the black dashed arrows in Fig. 3(a) and black solid arrows in Fig. 3(c). The radius 163nm corresponds to the magnetic dipolar resonance of a single particle [see Fig. 8 (a) in Appendix A].

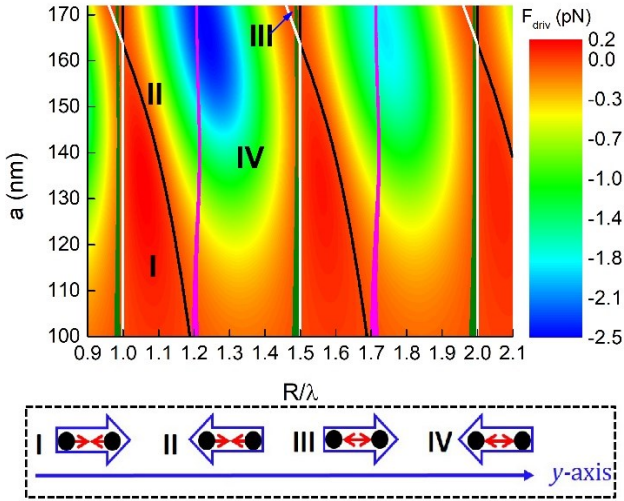

Fig. 4. Dependence of the driving force ($F_{driv}$) on the center of two polystyrene particles vs radius ($a$) and separation ($R$). The color denotes the magnitude of $F_{driv}$. The black (white) curves represent the stable (unstable) equilibrium positions of $F_{driv}$. The green (magenta) curves represent the stable (unstable) equilibrium separation of the two particles. The intersections of the white and black lines are located at $a$=163nm. Inside the dotted rectangle at the bottom of the figure, the black solid circles represent the particles. The thick blue arrows denote the moving direction along the $y$ or –$y$ axis (blue long directional ray) of the center of the dimer. The thin red arrows show the relative motion (attraction or repulsion) between the two particles in areas I to IV.

On the other hand, the green and magenta curves represent individually the stable and unstable equilibrium separations of the two particles leaded by the LOBFs. The two kinds of separations do not vary with the particle radius. This is different than those of the reciprocal $F_{Ray}$ whose unstable equilibrium positions gradually approach the fixed stable ones with increase of the particle size [see Fig. 9 (a) in Appendix B]. In term of dynamics, the positive $F_{driv}$ pushes the dimer along the $y$ axis (thick blue arrow) while the two particles attract each other (thin red arrows) in area I. However, the two attractive particles (thin red arrows) move along the -$y$ axis (thick blue arrow) in area II. On the contrary, the two particles repel each other in areas III and IV. But the whole dimer moves individually along the $y$ and -$y$ axes in the two regions. Moreover, the intensity of the negative $F_{driv}$ largely exceeds that of the positive one for wide separation and radius ranges, especially for large particles ($a$>130nm). It means that the large dimer favors the movement along the –$y$ axis. Finally, the directions of the motions in region 1.5$\lambda$ <$R$<2$\lambda$ repeat those in region $\lambda$<$R$<1.5$\lambda$ because the LOBF and $F_{driv}$ are periodic at $\lambda$/2. The phenomena are similar for silica particles with lower $n_p$=1.45, which are commonly used in optical binding experiments [37]. The difference is only that the intersections of the stable and the unstable equilibrium positions of $F_{driv}$ move to $a$=198nm.

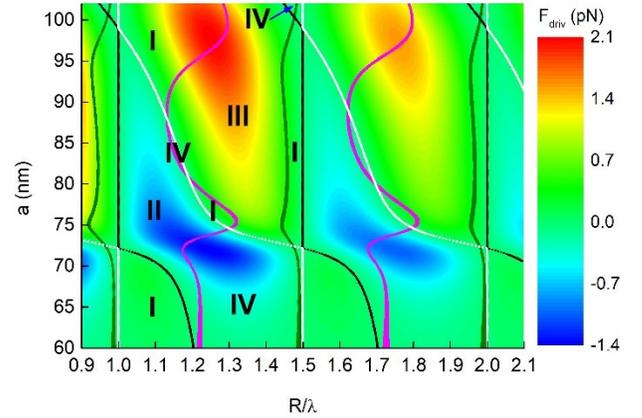

Fig. 5. Dependence of the driving force ($F_{driv}$) on the center of two silicon ($n_p$=3.5) particles vs radius ($a$) and separation ($R$). The meanings of the black, white, green, and magenta curves as well as the areas I to IV are same as the counterparts in Fig. 4.

Higher-refractive-index particles, e.g., silicon ($n_p$=3.5), have also been previously harnessed in optical micromanipulation [38, 39]. The reason is that their strong electric and magnetic responses [40, 41] greatly enhance the optical force [30, 42] compared to polystyrene particles. Figure 5 exhibits the much richer dynamics of the silicon homodimer and the two constituents, compared to the lower-refractive-index polystyrene particles in Fig. 4. The choice of the range (60nm-102nm) of the particle size is also based on the same reasoning as in Fig. 4 [see Fig. 8 (b) in Appendix A]. Firstly, the comparable intensities of the positive and negative maximum values of $F_{driv}$ means that the silicon homodimer can be forced to move along either the $y$ or –$y$ axes. For example, two particles with $a$=72nm in Fig. 6 (b) move together along the –$y$ axis while the collective movement of two 99nm particles in Fig. 6 (d) move along the $y$ axis regardless of their separation. Secondly, what is even more interesting than the polystyrene particles (Fig. 4) are that the stable (black curves) and unstable (white curves) equilibrium positions of $F_{driv}$ exchange multiply at positions $a$=72nm and 99nm. The two radii correspond respectively to the electric and magnetic dipolar resonances of a single particle [see Fig. 8 (b) in Appendix A]. For instance, the unstable equilibrium positions (black dash arrows) of $F_{driv}$ for 65nm particles in Fig. 6 (a) transform into the stable ones (black solid arrows) for 85nm particles in Fig. 6 (c).

On the other hand, the stable (green curves) and unstable (magenta curves) equilibrium separations between the two particles remain unchanged and change very little respectively when $a<67$nm. They are consistent with the counterparts of the $F_{Ray}$ [see Fig. 9 (b) in Appendix B] because the ED dominates over the MD [see Fig. 8 (b) in Appendix A] in this region. Interestingly, in the range of $a>67$nm, the formers fluctuate slightly on the left of the positions $R=l\lambda/2$, i.e., $\lambda$, $1.5\lambda$, and $2\lambda$, with particle size while the unstable ones fluctuate largely around $1.2\lambda$ and $1.7\lambda$. They are also shown by the green solid and dash arrows for different particle size in Fig. 6. In this range, both ED and MD resonances arise [see Fig. 8 (b) in Appendix A]. The two kinds of positions are different than the counterparts of the polystyrene particles in Fig. 4, which are independent of the particle size. Moreover, they are also distinct from those of $F_{Ray}$, where the unstable equilibrium positions move gradually towards the fixed stable ones with increase of the particle size [see Fig. 9 (b) in Appendix B]. In particular, the two particles attract each other in the region between the green (left side of $\lambda$) and magenta (around $1.2\lambda$) curves while they are repulsive in the neighboring region between the magenta (around $1.2\lambda$) and green (left side of $1.5\lambda$) curves. In addition, the direction of the motion of the dimer varies with the particle size and separation. In detail, the two particles attract each other in areas I and II. But the center of the dimer moves along the $y$ axis in area I since $F_{driv}$ is positive, while it moves along the $-y$ axis in area II due to the negative $F_{driv}$. On the other hand, the two particles repel each other in areas III and IV. But the center of the dimer moves along the $y$ axis in area III while it moves along the $-y$ axis in area IV. Moreover, the directions of the two kinds of movements are periodic in the two regions between $\lambda$ and $1.5\lambda$ and between $1.5\lambda$ and $2\lambda$.

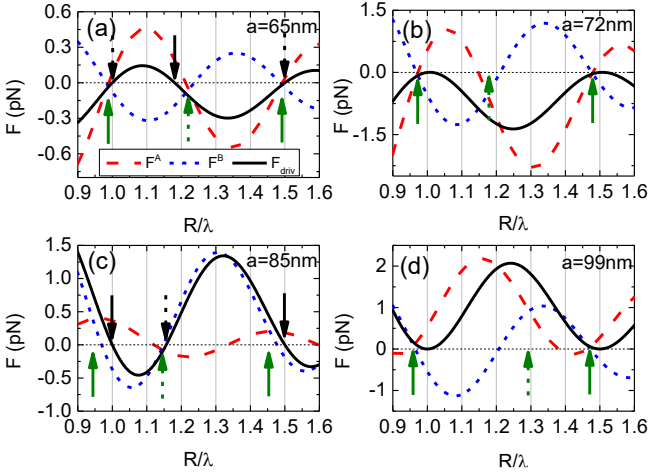

Fig. 6. LOBFs $F^A$ (red dashed curve) and $F^B$ (blue short dashed curve) on two silicon particles with radius $a=65$nm (a), 72nm (b), 85nm (c), and 99nm (d) as well as the driving force $F_{divi}$ (black solid curve) on the center of the dimer as a function of the separation ($R$). The black solid (dashed) arrows denote the stable (unstable) equilibrium positions of $F_{driv}$ while the green solid (dashed) arrows denote the stable (unstable) separations between the two particles.

## IV. DISCUSSION

The two ED moments and two MD moments distributions in a polystyrene homodimer with fixed $R=1.2\lambda$ are presented in Fig. 7 as functions of $a$ and $n_p$. The color and line type of the curves are the same as the counterparts in Fig. 1(b). Figure 7(a) shows that the ED pair is symmetric in the range $a<80$nm (Rayleigh region) because of the negligible magnetic dipolar response. Moreover, when $80$nm$<a<110$nm, the ED pair and MD pair are still symmetric because of the weak electromagnetic hybridization and coupling of the homodimer due to the small MD moments. As a result, LOBF is reciprocal for low-refractive-index and small particles ($a<110$nm), i.e., the typical parameters for widely-used silica and polystyrene particles. The nonreciprocal dynamics are hidden and the behavior is well-described by the Rayleigh approximation. This is also true for sub-wavelength metallic particles with only electric dipolar response. However, the asymmetric ED and MD pairs arise in the range $a>110$nm where the remarkable hybridization and coupling due to the strong electric and magnetic dipolar responses are large. Thus, these result in a nonreciprocal LOBF.

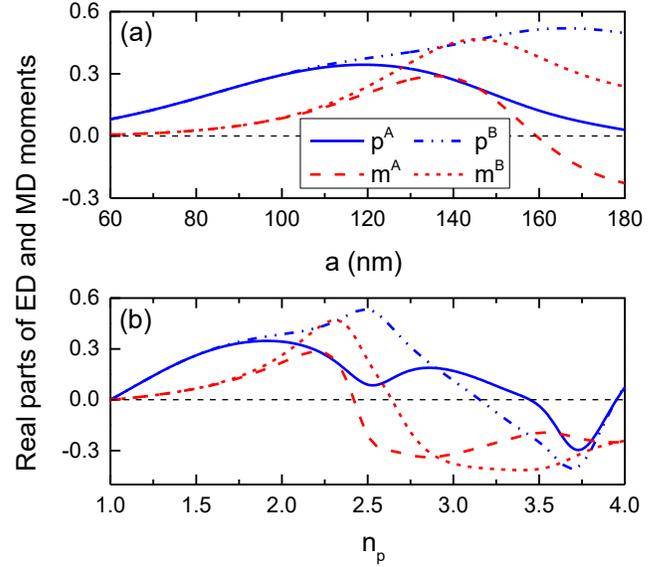

Fig. 7. The real parts of the dimensionless $p^A$, $p^B$, $m^A$, and $m^B$ in a dielectric homodimer as function of radius $a$ (a) and refractive index $n_p$ (b). The separation is fixed at $R=1.2\lambda$ and the parameters of the incident wave are the same as those in Fig. 2 while $n_p=1.59$ (a) and $a=100$nm (b).

On the other hand, Fig. 7 (b) shows the four dipole moments in dielectric particles with $a=100$nm as function of $n_p$. It is clear that not only the ED pair but also the MD pair are symmetric when $n_p<1.8$ can be due to either the negligible magnetic dipolar response of the particles or the weak hybridization and coupling. Therefore, the nonreciprocal LOBF is hard to detect for particles with $n_p<1.8$ and $a=100$nm. On the contrary, the nonreciprocal LOBF arise when $n_p>2$, i.e., silicon, because of the obviously asymmetric ED pair and MD pair. Note that the shapes of the four curves in Fig. 7 (a) are basically the same as the shapes of the counterparts in Fig. 7 (b) in the range $1<n_p<2.5$. The reason is that the excitations of the ED and MD are determined by the size parameter $x=n_p ka$, which depends on the refractive index and radius of particle when the incident wavelength is fixed [43]. Therefore, the similar $x$ intervals from 1.1 to 3.4 in Fig. 7 (a) and from 1.2 to 3.0 where $1<n_p<2.5$ in Fig. 7 (b) result in the similar shapes of the curves. It means that the manipulations of the dipolar response of the particle can be realized by the variation of either $n_p$ or $a$. In other words, the realization of the nonreciprocal LOBF of dielectric homodimer can be implemented by choosing either large or high-refractive-index particles (termed Mie particle).

As expected, the electric and magnetic multipole moments, i.e., quadrupole, octupole, and so on, will be induced with increase of the particle size. Therefore, the anisotropic radiation fields of the multipolar moments with different phases may also contribute to the nonreciprocal LOBF as discussed in Appendix C and Fig. 10. However, the corresponding analytical study is beyond the scope of this paper. In addition, for micron-scale particles ($a>>\lambda$), optical binding is treated with geometrical optics [44]. Therefore, the LOBF at this scale is still reciprocal because of the reflection symmetry between the two identical particles. In summary, LOBF is nonreciprocal for a submicron polystyrene homodimer with $a>120$nm or 100nm dielectric particles with $n_p >2$.

## V. CONCLUSIONS

In summary, we have presented the analytical expressions of the LOBF between two identical electric and magnetic dipolar dielectric particles. It is composed of the electric dipolar, magnetic dipolar, and electric-magnetic dipolar coupling interactions. Our results show that not only the LOBF has been obviously underestimated but also the stable and unstable equilibrium positions of the force cannot be accurately predicted in Rayleigh approximation in the circumstances where the particle size or refractive index are large, e.g., for polystyrene particles when the radius of the particles are above 130nm. The reasons are that the contributions of the magnetic and coupling interactions on the force are ignored in the Rayleigh limit. Surprisingly, the LOBF is asymmetric for the case of two identical particles. This is a result of the symmetry-breaking of the forward and backward scattering of the particles due to the electric-magnetic dipolar coupling interaction. Such a difference could be visualized experimentally in optical binding experiments using ultrashort pulsed lasers [45]. In addition, the resulting driving force on the center of the homodimer is harmonic and decays with the interparticle distance to the first power. Remarkably, the stable and unstable equilibrium positions of the driving force exchange depending on the particle size and interparticle distance. They are contrary to those of the LOBF. Therefore, both the driving force and asymmetric LOBF lead to rich nonequilibrium dynamics of the dimer and of the two constituents. Finally, we have presented the refractive index and size ranges of dielectric particle required for the asymmetric LOBF.

Although the asymmetric binding force has been previously demonstrated to exist between mismatched particles [18, 24], the reason for the effect here is completely different than that described in that study. Moreover, the asymmetric LOBF requires no additional phase gradient which causes the asymmetric TOBF for two identical particles in an optical ring trap [22]. Our results provide new insight into nonreciprocal optical binding and open perspectives for nonequilibrium dynamics of light-driven nanomotors and multiparticle self-assembly.

## ACKNOWLEDGMENTS

This work was supported by Jiaxing Science and Technology Project Grants (No.2021AY10057), KD acknowledges support of the UK Engineering and Physical Sciences Research Council (grant EP/P030017/1).

## APPENDIX A: ELECTRIC AND MAGNETIC MULTIPOLAR COMPONENTS OF SCATTERING COEFFICIENTS

The LOBF and driving force come from the light scattering from the particles. Therefore, the proportions of the electric and magnetic multipolar components in the scattering coefficient ($Q_{sca}$) represent the weights of the electric and magnetic multipolar interactions in LOBF and driving force. In this section, the contributions of the electric and magnetic multipole moments on $Q_{sca}$ are discussed. $Q_{sca}$ is defined as the ratio of the scattering cross-section to the cross-section area of the particle [43]. Figure 8 shows the $Q_{sca}$ including its electric and magnetic dipolar (ED and MD), quadrupolar (MQ and MQ), and octupolar (EO and MO) components of single polystyrene (a) and silicon (b) particles. For the polystyrene particle in Fig. 8 (a), the $Q_{sca}$ (green dash-doted curve) is dominated by the ED (black solid curve) when $a<100$nm. Thus, the particle can be regarded as a Rayleigh particle in the size range. However, the ED and MD (black dashed curve) are effectively excited and determine together the $Q_{sca}$ in the range from 100nm to 172nm. Particularly, the MD resonance arises at $a=163$nm while the ED is non-resonant. Furthermore, the EQ (red solid curve) and MQ (red dashed curve) are effectively excited in region 172nm<$a$<250nm while the excitations of the EO (blue solid curve) and MO (blue dashed curve) are inconspicuous. Hence, the particles in this size region go beyond our electric and magnetic dipolar model. On the other hand, for the silicon particle in Fig. 8 (b), the ED almost determines $Q_{sca}$ when $a<67$nm. Furthermore, The ED and MD dominate together $Q_{sca}$ in the range 67nm<$a$<102nm within our electric and magnetic dipolar model region. However, the particles with $a>102$nm have strong MQ resonance and are out of our model range.

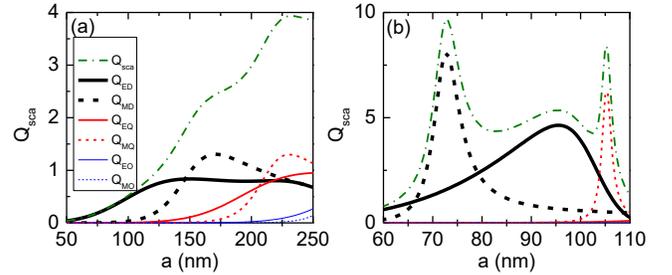

Fig. 8. The scattering coefficients $Q_{sca}$ (green dash-doted curve) including its electric and magnetic multipolar components of single polystyrene (a) and silicon (b) particles as a function of the particle radius ($a$). The thick black solid (dashed) curve denotes the electric (magnetic) dipolar component. The red solid (dashed) denotes the electric (magnetic) quadrupolar component. The thin blue solid (dashed) denotes the electric (magnetic) octupolar component.

## APPENDIX B: $F_{RAY}$ AND ITS STABLE AND UNSTABLE EQUILIBRIUM POSITIONS

For Rayleigh particles, e.g., polystyrene (50nm<$a$<130nm) and silicon (30nm<$a$<65nm) spheres, the classical LOBF $F_{Ray}$ are respectively shown in Figs. 9 (a) and 9 (b). The color represents the intensity of the force. The green and magenta curves show individually the stable and unstable equilibrium positions of $F_{\text{Ray}}$. It can be seen that the stable equilibrium positions remain unchanged while the unstable ones approach slowly the stable ones with increase of the particle radius. However, the two kinds of the positions do not intersect each other in our parameter ranges.

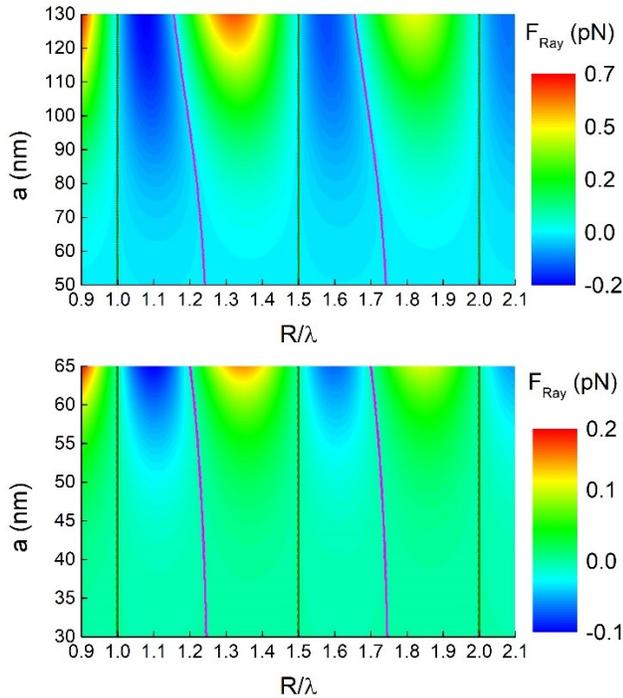

Fig. 9. Dependence of $F_{\text{Ray}}$ between two polystyrene particles (a) and two silicon particles (b) vs radius ($a$) and separation ($R$). The color denotes the magnitude of the force. The green and magenta curves represent respectively the stable and unstable equilibrium positions of $F_{\text{Ray}}$.

## APPENDIX C: THE EFFECTS OF THE MULTIPOLES IN PARTICLES ON THE ASYMMETRIC BINDING FORCE

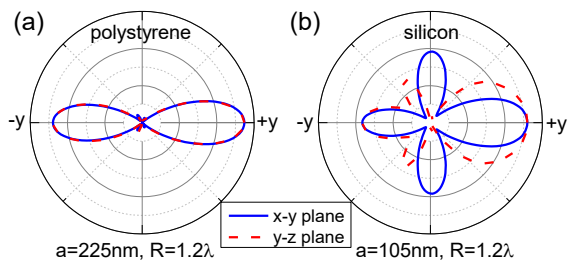

Fig. 10. The far-field scattering patterns of a polystyrene homodimer with $a$=225nm and $R$=1.2λ including the electric and magnetic quadrupoles (a) and a silicon homodimer with $a$=105nm and $R$=1.2λ including a dominant magnetic quadrupole (b) in $x$-$y$ (blue solid curves) and $y$-$z$ (red dash curves) planes.

In this section, we show the effects of the electric and magnetic quadrupoles on the asymmetric LOBF. For instance, the far-field scattering pattern of two polystyrene particles with $a$=225nm and $R$=1.2λ is presented in Fig. 10 (a). The forward scattering dominating the backward scattering means that the dimer experiences a net optical driving force directing in the –$y$ axis. Meanwhile, this demonstrates an asymmetric LOBF between the two particles. The asymmetry of the binding force has contributions due to the asymmetric radiation coming from not only the dipoles but also the non-negligible electric and magnetic quadrupoles [see Fig. 8(a)]. The analogous phenomenon occurs for two silicon particles with $a$=105nm and $R$=1.2λ in Fig. 10 (b) where the particles have dominant magnetic quadrupole resonance as shown in Fig. 8(b). Therefore, the appearances of the electric and magnetic quadrupoles also contribute to the asymmetric LOBF. In these regimes, an even richer physics is expected, but the analytical investigation of this is beyond the scope of this present study.